\documentstyle[prl,aps,twoside,floats,epsf]{revtex}
\begin{document}
\draft
\twocolumn[\hsize\textwidth\columnwidth\hsize\csname
@twocolumnfalse\endcsname
\begin{title}
{The  essential interactions in oxides and spectral weight transfer 
 in doped manganites}
\end{title} 
\author{ A.S. Alexandrov$^{1,*}$ and A.M. Bratkovsky$^{2,\dagger}$}
\address{$^{1}$Department of Physics, Loughborough University,
Loughborough LE11 3TU, UK\\
$^{2}$Hewlett-Packard Laboratories, 3500~Deer~Creek
Road, Palo Alto, California 94304-1392 }
\date{August 11, 1999}
\maketitle
\begin{abstract}
We calculate the value of the Fr\"ohlich electron-phonon interaction
in manganites, cuprates,  and some other charge-transfer insulators and
show that this interaction is much stronger than any relevant magnetic
interaction.
A polaron shift due to the Fr\"ohlich interaction, which is about 1~eV,
suggests that carriers in those systems are small (bi)polarons at all
temperatures and doping levels, in agreement with the oxygen isotope
effect and  other data.  
An opposite conclusion, recently suggested in the literature, 
is shown to be incorrect. The frequency and temperature
dependence of the optical conductivity of ferromagnetic manganites
is explained within the framework of the bipolaron theory.

\pacs{71.38.+i, 74.20.Mn, 75.30.Vn, 78.20.-e}
\end{abstract}

\vskip 2pc ] 
\narrowtext

\section{Introduction}

It is well understood that carriers in cuprates\cite{alemot,mul2} and
manganites\cite{mil,bis,alebra} are strongly 
coupled to lattice vibrations. We have recently proposed
a bipolaron theory of ferromagnetism and colossal magnetoresistance
(CMR) based on the idea of a current carrier density collapse (CCDC) due to
an interplay of the electron-phonon and exchange interactions in doped
manganites \cite{alebra}. Owing to the strong electron-phonon interaction,
polaronic carriers are bound into almost immobile bipolarons in the
paramagnetic phase of CMR materials. The non-degenerate
polarons induce a polarization of localized Mn $d$ electrons. As a result,
the exchange interaction dissociates bipolarons below $T_{c}$ if the $p-d$
exchange energy $J_{pd}S$ of the polaronic carriers with the localized Mn $d$
electrons is larger than the bipolaron binding energy $\Delta $. Hence, the
density of current carriers (polarons) suddenly increases below $T_{c}$,
which explains the resistivity peak and CMR, observed in many ferromagnetic
oxides \cite{van,helmolt,jin}. We have shown \cite{alebra2} that CCDC also
explains the giant isotope effect \cite{mul,isoJF}, the tunneling gap
\cite{tun}, 
the specific heat anomaly \cite{ram}, and the temperature dependence of the
dc resistivity \cite{sch}.

The non-metallic nature of the ferromagnetic low-temperature phase of the
doped manganites has been unambiguously confirmed in recent studies of the
optical conductivity \cite{oki1,oki2,kim,ish} and photoemission \cite{des}.
In particular, a broad incoherent spectral feature \cite{oki1,oki2,kim,ish}
in the midinfrared region and a pseudogap in the excitation spectrum \cite
{des} were observed, while the coherent Drude weight appeared to be { one
or even two orders} of magnitude smaller \cite{oki2,kim} than expected for a
metal, or almost {\em absent}\cite{ish}. These and other studies \cite{gud}
prove that carriers retain their polaronic character well below $T_{c}$, in
agreement with our theory of CMR \cite{alebra}. For example, the measured
residual conductivity (at $T=0$) $\sigma _{0}=360\Omega ^{-1}$cm$^{-1}$\cite
{ish} yields the product of the Fermi wave vector and the mean free path $%
k_{F}l\lesssim 1$, which is below the Ioffe-Regel limit\cite{sigma}.
Hence,
Fermi-liquid type theories appear to be inadequate even for a description of
the low-temperature phase of manganites.

Among the major phenomena yet to be explained in the manganites are the
spectral weight transfer with temperature \cite{oki2,kim,ish} and the
pronounced peak structure \cite{ish} in the optical conductivity both above
and below $T_c$, observed in several systems.

In this paper we first calculate the value of the electron-phonon
interaction in oxides. We then propose a theory of the optical conductivity,
including the {\em massive spectral weight transfer} below the ferromagnetic
transition, based on the idea of the current carrier density collapse. We
show that the high-temperature optical conductivity is well described by the
small bipolaron absorption, while the low temperature midinfrared band is
due to absorption by small polarons. The bipolaron dissociation below $T_c$
shifts the spectral weight from the bipolaronic peak to the polaronic one.
We describe the optical spectra of the layered ferromagnetic ($T_c=125$K)
crystal La$_{2-2x}$Sr$_{1+2x}$Mn$_2$O$_7$ \cite{ish} in the entire frequency
and temperature range that has been studied experimentally, and show that
the optical data provides strong evidence for CCDC.

\section{Fr\"ohlich interaction and small (bi)polaron formation in oxides}

The carriers in both cuprates and manganites are O~$p$ ({\em not} Cu or Mn~$d$)
holes \cite{kannan}, and their pairing in the paramagnetic phase due to
strong electron-phonon interaction would most likely be in the form of {\em %
inter}site bipolarons localized on a pair of O sites \cite{alemot}. Clearly,
the formation of these bipolarons cannot be hindered by repulsive
on-site Hubbard-$U$ and/or Hund's rule coupling, contrary to recent claims
in the literature \cite{millis98}.

In order to assess the possibility of small polaron and bipolaron
formation in manganites and cuprates, one has to calculate the
electron-phonon interaction and compare it with the (inter-site) Coulomb
repulsion. Fortunately, such an estimate is possible in oxides, which are
highly polarizable materials with a substantial long-range Fr\"{o}hlich
interaction. Then the polaron binding energy $E_{p}$ (polaronic shift) can
be explicitly expressed through the well defined experimental parameters. In
the long-wave limit the response of polarons at the optical phonon frequency
is dynamic \cite{aleres}. Their (renormalized) plasma frequency is lower
than the optical phonon frequency due to the large static dielectric constant
of oxides, an enhanced effective mass, and relatively low density of
polarons. Therefore, the long-range character of the Fr\"{o}hlich
interaction is unaffected by screening. The optical phonon frequency
is hardly affected as well \cite{alesch}.  Claims in the literature \cite
{cha,cha2} that the Fr\"{o}hlich interaction is reduced to a local Holstein
interaction in doped oxides disregard the established fact that the 
{ mobility}
of carriers { determines the screening}, rather than their
density\cite{alemot}. 

We shall apply the exact expression for the polaronic shift $E_{p}$ \cite
{yam,lan,eag,alemot}, which for the Fr\"{o}hlich interaction reads 
\begin{equation}
E_{p}={\frac{1}{{2\kappa }}}\int_{BZ}{\frac{d^{3}q}{{(2\pi )^{3}}}}{\frac{%
4\pi e^{2}}{{q^{2}}}}.
\label{eq:epF}
\end{equation}
Here the dielectric constants, 
$\kappa^{-1}=\epsilon_{\infty
}^{-1}-\epsilon _{0}^{-1}$, 
are known from experiment, and the size of the
integration region (the Brillouin zone, BZ) is determined by the lattice
constants $a,b,c$ (Table). The data in the Table represent the {\em lower}
boundary for the polaron binding energies, Eq.~(\ref{eq:epF}), since 
the account for coupling with acoustic and/or
Jahn-Teller modes can only increase the polaronic shift $E_p$.
The correct value of $E_p$
appears to be about $1${\rm eV} in most cases, far exceeding
the ad hoc estimates of $\sim 0.15$ {\rm eV}\cite{nag}, obtained from a
numerically incorrect expression for $E_{p}$ and wrong values of the
dielectric constants.
 The large calculated value of the
 polaron shift ($\sim 1$eV) is perfectly
 compatible with the small polaron theory.
According to \cite{nag}  the bare
half-bandwidth is about $W/2\simeq 0.8$ eV 
in manganites. Hence, even a simple variational criteria
\cite{nag} of the small polaron formation ($E_{p}>W/2$) is satisfied.
In the case of the Fr\"ohlich interaction the
small polaron theory based on the Lang-Firsov canonical transformation is
numerically accurate even in the weak coupling (large polaron) regime
irrespective of the value of $E_{p}$\cite{alekor}.
 It is much more important, irrespective of any theoretical arguments, 
 that the existence of polarons in CMR materials is unambiguosly
 confirmed experimentally, including: very low mobility\cite{sch,ish}, 
 incompatible with Boltzmann-type approaches, 
activated dc  and ac transport in the paramagnetic phase\cite{ish,jaime}  and
the giant isotope effect in manganites\cite{mul,isoJF}.

 \begin{table}[ht]
 \caption{
 Polaron shift $E_{p}$ due to Fr\"ohlich interaction. Data
 from {\it Handbook } {\it of Optical} {\it  Constants} {\it of
 Solids}, edited by E.D. Palik (Academic, New York, 1997) and {\it
 Landolt}-{\it B\"ornstein, Group III}. The value $\epsilon _{\infty
 }=5$ for 
 WO$_3$ is an estimate.  }
 \begin{tabular}{lllll}
 System & $\epsilon _{\infty }$ & $\epsilon _{0}$ & $a\times b\times c$ (\AA $%
 ^{3}$) & E$_{p}$(eV) \\ \hline
 BaBiO$_{3}$ & 5.7\tablenotemark[1]
 & 30.4\tablenotemark[2] 
 & 4.34$^{2}\times $4.32 & 0.579 \\ 
 BaTiO$_{3}$ & 5.1-5.3 & 1499. & 3.992$^{2}\times $4.032 & {0.842} \\ 
 La$_{2}$CuO$_{4}$ & 5.0 & 30 & 3.8$^{2}\times
 $6\tablenotemark[3]
 &   {0.647} \\ 
 LaMnO$_{3}$ & 3.9\tablenotemark[4]
 & 16\tablenotemark[4]\tablenotemark[5]
 & 3.86$^{3}$ &   {0.884}
 \\ 
 La$_{2-2x}$Sr$_{1+2x}$ &  &  &  &   {} \\ 
 $-$Mn$_{2}$O$_{7}$ & 4.9\tablenotemark[4]
 & 38\tablenotemark[4] 
 & 3.86$^{2}\times $3.9\tablenotemark[6]
 &   {0.807} \\ 
 SrTiO$_{3}$ & 5.2 & 310 & 3.905$^{3}$ &   {0.852} \\ 
 WO$_{3}$ & 5 & 100-300 & 7.31$\times $7.54$\times $7.7 & {0.445} \\ 
 CdO & 5.4 & 21.9 & 4.7$^{3}$ &   {0.522} \\ 
 EuS & 5.0 & 11.1 & 5.968$^{3}$ &   {0.324} \\ 
 EuSe & 5.0 & 9.4 & 6.1936$^{3}$ &   {0.266} \\ 
 MgO & 2.964 & 9.816 & 4.2147$^{3}$ &   {0.982} \\ 
 NaCl & 2.44 & 5.90 & 5.643$^{3}$ &   {0.749} \\ 
 NiO & 5.4 & 12 & 4.18$^{3}$ &   {0.429} \\ 
 TiO$_{2}$ & 6-7.2 & 89-173 & 4.59$^{2}\times $2.96 &   {0.643
 } \\ 
 \end{tabular}
 \tablenotetext[1]{S. Uchida {\it et al.} J. Phys. Soc. Jap. {\bf 54},
 4395 (1985).}
 \tablenotetext[2]{R.P.S.M. Lobo and F. Gervais, Phys. Rev. B {\bf 52}, 13294
 (1995).}
 \tablenotetext[3]{Distance between CuO$_2$ planes.}
 \tablenotetext[4]{T. Ishikawa (private communication, 1999).}
\tablenotetext[5]{J.H. Jung and T.W. Noh (private communication, 1999)
have estimated from their data  $\epsilon_\infty=3.4$ and $\epsilon_0=21$
for LaMnO$_3$.}
 \tablenotetext[6]{Distance between MnO$_{2}$ planes.}
 \end{table}

The effective polaron-polaron attraction, due to the overlap of the
deformation fields, is about $2E_{p}$ \cite{alemot}. This appears to
be more than 
sufficient to overcome the intersite Coulomb repulsion, $V_{c}\simeq 0.7-0.8
{\rm eV}$, as also confirmed by the first-principles lattice minimization
technique \cite{cat}. Hence, we can conclude that the electron-phonon
interaction is comparable or even  stronger than any relevant 
magnetic interaction
(which is
estimated as $\lesssim 0.2$ eV \cite{nag}), so that small (bi)polarons are
indeed the most probable quasiparticles in oxides, in contrast to the
erroneous claims \cite{nag}.

\section{Optical conductivity of intersite bipolarons}

Now  we  show how the observed {\em massive} weight transfer in optical
conductivity with temperature\cite{ish} can be naturally explained within
the bipolaron theory. The optical intraband conductivity of a
charge-transfer doped insulator with (bi)polaronic carriers is the sum of
the polaron $\sigma _{p}(\nu )$ and bipolaron $\sigma _{b}(\nu )$
contributions at a given frequency $\nu $. Their frequency dependences are
described in the literature \cite{bri,mah,alemot,kli,eag2,rei,bri2}. In the
leading saddle-point approximation both have almost a Gaussian shape given
by 
\begin{eqnarray}
\sigma _{{\rm intra}}(\nu ) &=&{\frac{\sigma _{0}{\cal T}^{2}}{{\nu }}}%
\Biggl[{\frac{n}{{\ \gamma _{p}}}}\exp \left[ -(\nu -\nu _{p})^{2}/\gamma
_{p}^{2}\right]  \nonumber \\
&+&{\frac{x-n}{{\gamma _{b}}}}\exp \left[ -(\nu -\nu _{b})^{2}/\gamma
_{b}^{2}\right] \Biggr],
\label{eq:sintra}
\end{eqnarray}
where $\sigma _{0}=2\pi ^{1/2}e^{2}/a$ is a constant with $a$ the lattice
spacing, ${\cal T}$ the hopping integral, $n$ the (atomic) polaron density,
and $x$ the doping level. Here and further we take $\hbar =c=1$.

We shall first determine the positions and widths of the (bi)polaron
absorption peaks for oxides. Those are known for the Holstein model with
local interactions, but it is unlikely to apply for oxides because of very
large on-site Coulomb repulsion and the long-range (Fr\"{o}hlich)
electron-phonon interaction, which dominates in ionic solids.

\begin{figure}[t]
\epsfxsize=2.4in
\epsffile{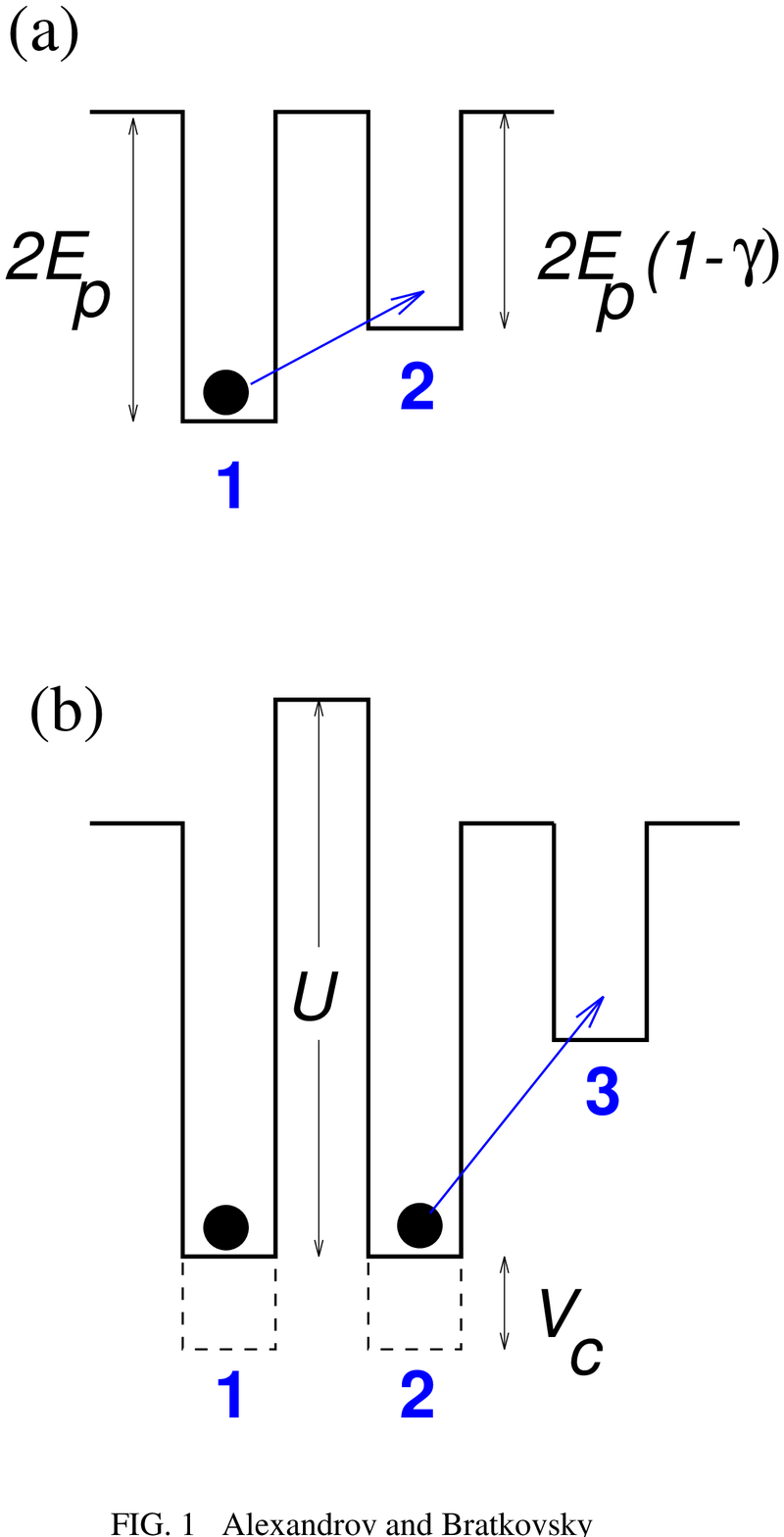 }
\vspace{.2in}
\caption{Adiabatic energy levels and optical trannsitions for the
small polaron (a)
and the inter-site small bipolaron (b).
\label{fig:1}
}
\end{figure}
Applying the Franck-Condon principle\cite{mah} in the adiabatic regime, $%
\nu\gg\omega$ (phonon frequency), one can generalize the (bi)polaronic
absorption, 
Eq.~(\ref{eq:sintra}), to 
describe the optical conductivity of small Fr\"ohlich (bi)polarons, Fig.~1. The
electron `sitting' on a site ``1'', Fig.~1(a), lowers its energy by an
amount $2E_{p}$, with respect to an atomic level in the undeformed lattice,
owing to the lattice deformation. If the electron-phonon interaction has a
finite radius, the electron also creates some deformation around a
neighboring site ``2'', lowering its energy level by an amount $%
2E_{p}(1-\gamma)$, where \cite{ale}: 
\begin{equation}
\gamma=\sum_{{\bf q}}|\gamma({\bf q})|^2\left[1-\cos({\bf q \cdot a})\right]
/\sum_{{\bf q}}|\gamma({\bf q})|^2
\label{eq:gamma}
\end{equation}
with ${\bf a}$ the lattice vector connecting the neighboring sites. The
coefficient $\gamma$ strongly depends on the radius of the interaction. In
the Holstein model with ${\bf q}$-independent electron-phonon coupling, $%
\gamma({\bf q})$, this coefficient equals unity. Hence, there is no lattice
deformation at the neighboring site. On the contrary, in the Fr\"ohlich
case, $\gamma({\bf q})\propto 1/q$, and the coefficient is quite small, $%
\gamma \approx 0.2-0.4$ \cite{ale} depending on the dimensionality of the
system and the unit cell geometry. In that case, there is a significant
lowering of the neighboring energy level and, as a result, of the polaron
mass \cite{alekor}. Hence, generally, the peak energy in the polaron
absorption is found at 
\begin{equation}
\nu_{p}= 2 \gamma E_p,
\label{eq:nup}
\end{equation}
and the activation energy of the high-temperature dc-conductivity is $%
E_a=\gamma E_p/2$\cite{mah}. One can apply the same `frozen lattice
distortion' arguments to the inter-site bipolaron absorption, Fig.~1(b). The
electron energy on site ``2'' is $-2E_{p} -2E_{p}(1-\gamma)+ V_{c}$, where
the first contribution is due to the lattice deformation created by the
electron itself, while the second contribution is due to the lattice
deformation around the site ``2'' created by the other electron of the pair
on the site ``1'', which is the polaron-polaron attraction \cite{alemot}.
After absorbing a quantum of radiation, the electron hops from site ``2'' to
the empty site ``3'' into a state with the energy $-2E_{p}(1-\gamma)$, which
corresponds to an absorption frequency 
\begin{equation}
\nu_{b}= 2E_{p}-V_{c},  
\label{eq:nub}
\end{equation}
where $V_{c}$ is the inter-site Coulomb repulsion. The quantum broadening of
the polaronic and bipolaronic absorption is given by $\gamma_p=\gamma_b=
(4\gamma E_p\omega)^{1/2}$. Since doped manganites are intrinsically
disordered, their dielectric properties are inhomogeneous, and so is $E_p$,
which fluctuates with a characteristic impurity broadening $\Gamma_{{\rm im}%
} $. The convolution of the polaronic and bipolaronic absorption lines with
the Gaussian distribution of $E_p$ results in their having different
linewidths, $\gamma_p= 2(\gamma E_p\omega +\gamma^2\Gamma_{{\rm im}}^2
)^{1/2}$ and $\gamma_b=2(\gamma E_p\omega+\Gamma_{{\rm im}}^2 )^{1/2}$ for
polaronic and bipolaronic absorption, respectively. The Coulomb repulsion $%
V_c$ can be readily estimated as $V_{c}=2E_p - \nu_b$ from (\ref{eq:nub}).

\section{\ Spectral weight transfer in manganites}

The total absorption is the sum of the intraband polaronic and bipolaronic
terms, Eq.~(\ref{eq:sintra}), and the interband absorption, $\sigma
(\nu )=\sigma _{{\rm %
intra}}(\nu )+\sigma _{{\rm inter}}(\nu )$. In the layered compounds like La$%
_{2-2x}$Sr$_{1+2x}$Mn$_{2}$O$_{7}$, the intraband contribution to the
out-of-plane conductivity is negligible \cite{ish}. Hence, one can take the
c-axis optical conductivity $\sigma _{c}(\nu )$ as a measure of the
interband contribution to the in-plane conductivity with a scaling factor, $%
s $, $\sigma _{{\rm inter}}(\nu )\simeq s\sigma _{c}(\nu )$. The scaling
factor $s$ is the square of the ratio of the in-plane components of the
dipole matrix element for the interband transitions to its $z$ component ($z$
is the out-of-plane direction). It can be readily determined by comparing
the in-plane and out-of-plane optical conductivities at high frequencies,
where intraband absorption is irrelevant. The result of the comparison of
the present theory with the experiment \cite{ish} is shown in Fig.~2. 
\begin{figure}[t]
\epsfxsize=3.4in
\epsffile{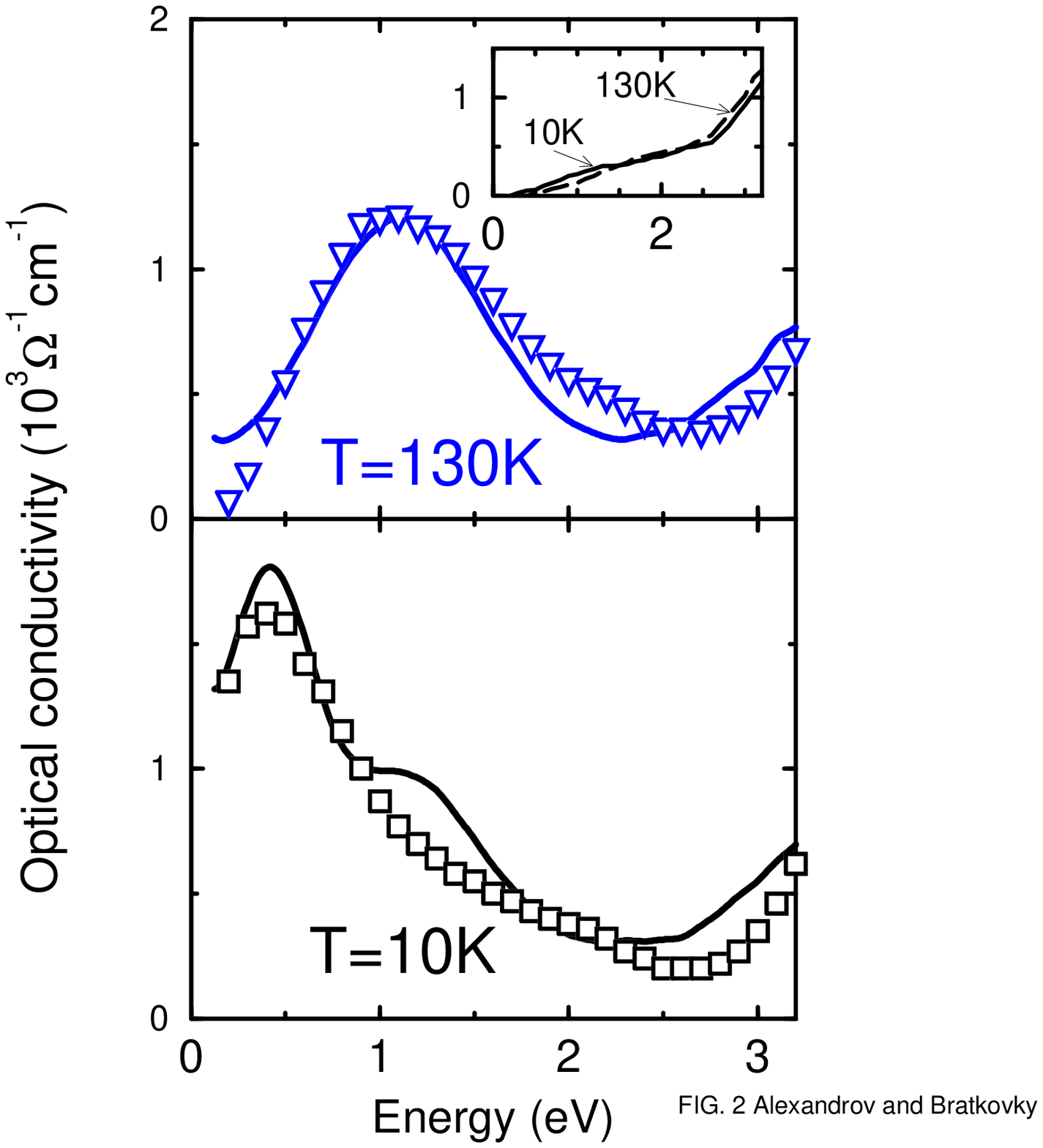 }
\vspace{.2in}
\caption{Optical conductivity of La$_{2-2x}$Sr$_{1+2x}$Mn$_2$O$_7$
[17] compared with the theory (solid line) above $T_{c}$ (top panel) and
well below
$T_{c}$ (bottom panel). Inset: experimental c-axis optical
conductivity [17].
\label{fig:2}
}
\end{figure}
At temperatures above the transition ($T=130$K) the polaron density is
very low owing to CCDC \cite{alebra}, so the intraband conductivity is
due to bipolarons only, 
\begin{equation}
\sigma (\nu )={\frac{\sigma _{0}x{\cal T}^{2}}{{\nu \gamma _{b}}}}\exp
\left[ -(\nu -\nu _{b})^{2}/\gamma _{b}^{2}\right] +s\sigma _{c}(\nu ).
\label{eq:shT}
\end{equation}

This expression fits the experiment fairly well with $\nu _{b}~=~1.24$ eV
and $\gamma _{b}~=~0.6$ eV, Fig.~2. The scaling factor is estimated as $%
s=0.6 $. When the temperature drops below $T_{c}$, at least some of the
bipolarons break apart by the exchange interaction with Mn sites, because
one of the spin-polarized polaron bands falls suddenly below the bipolaron
level by an amount $(J_{pd}S-\Delta )/2$, Fig.~3 \cite{alebra}. The
intraband optical conductivity is determined now by both the polaronic and
bipolaronic contributions, Eq.~(2), and that explains the sudden spectral
weight transfer from $\nu =\nu _{b}$ to $\nu =\nu _{p},$ observed below $%
T_{c}$ in the ferromagnetic manganites \cite{oki2,kim,ish}. The experimental
spectral shape at $T=10$K is well described by Eq.~(1) with $n=x/5$, $\nu
_{p}\simeq 0.5$ eV and $\gamma _{p}\simeq 0.3$ eV (Fig.~2). Taking $E_{p}=1$%
eV (Table) we find from Eq.(4) $\gamma \simeq 0.25$ corresponding to the
activation energy 
$E_{a}=125$meV and phonon frequency $\omega\simeq
75-90$meV\cite{romero}. With the use of Eq.(5) we find 
$V_{c}\simeq 0.76$eV in good correspondence with previous estimates for
cuprates\cite{ale}.

We note the weak temperature dependence of the optical conductivity at low
temperatures $T<50$K and above $T_{c}$ \cite{ish} in agreement with our
theory. We do not expect any significant temperature dependence of the optical
conductivity in the paramagnetic phase because the polaron density remains
small compared with the bipolaron density above $T_{c}$ \cite{alebra}.
Our results also suggest a {double-peak} structure of $\sigma
(\nu )$ below $T_{c}$, which was clearly observed for La$_{0.7}$Ca$_{0.3}$MnO%
$_{3}$\cite{noh}. This structure is not evident in the experiment on layered
compound (Fig.~2), thus, more detailed studies of this low-temperature
region are most desirable. It is worth mentioning that the scatter of the
on-site energies of carriers may lead to an asymmetric broadening of both
polaronic and bipolaronic absorption peaks\cite{BrVolRai83}, which can wash
out the double-peak structure at low temperatures.

 The temperature dependence of
the polaron density  can be found from our Hartree-Fock
equations \cite{alebra,alebra3}. 
The polaron density at zero temperature is about the total (chemical)
density of holes introduced by doping, 
while in the paramagnetic phase  one obtains 
\begin{equation}
n(T>T_{c})\sim \exp (-\Delta /2k_{B}T)
\end{equation}
for $T\geq T_c$ with the prefactor depending on the ratio
$W_p/T_c$. Note that $n\sim x$ below the Curie point, at 
$T<T_c$. Therefore, there is a {\em very}
abrupt drop of the polaron 
density at $T_{c}$, which {\em exponentially} depends on the bipolaron
binding energy $\Delta=\nu_{b}-\nu_{p}-W_{p}$ (Fig.~3). It can reach
many orders of magnitude if the relative strength of the exchange
interaction $J_{pd}S$ is close enough to  the binding energy.  
\begin{figure}[t]
\epsfxsize=3.4in
\vspace{.2in}
\epsffile{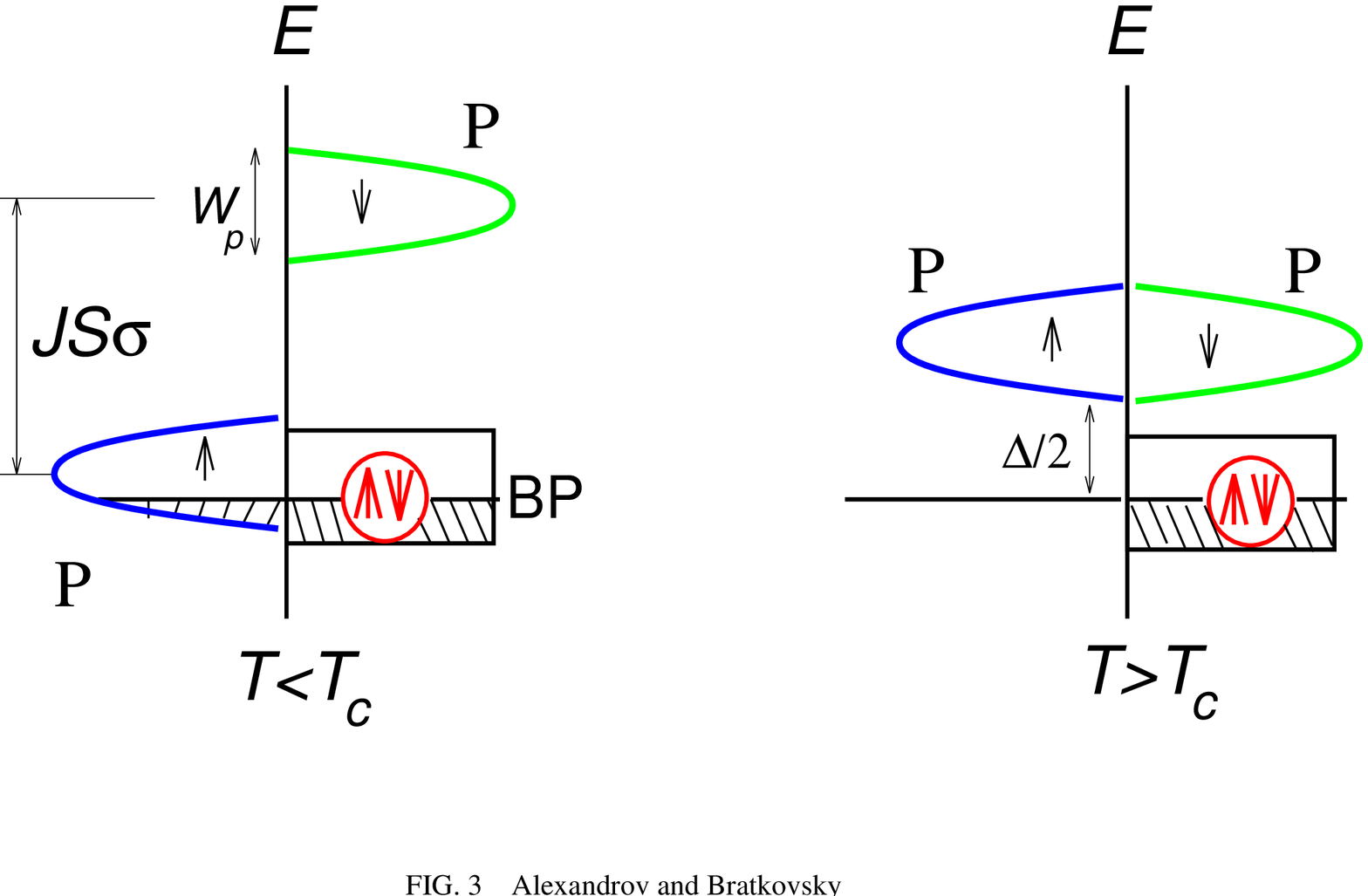}
\vspace{.2in}
\caption{Spin-polarized polaron band (P) in the ferromagnetic phase ($T<T_c$)
overlaps with the bipolaron (impurity) band (BP), breaking up
a fraction of the bipolarons. In the paramagnetic phase ($T>T_c$) the
spin-splitting of the polaron band is absent and system contains
mainly localized bipolarons.
\label{fig:3}
}
\end{figure}

\section{ Conclusions}

In conclusion, we have calculated the value of the electron-phonon
interaction in oxides (Table) to show that the Fr\"{o}hlich interaction
plays the dominant role in comparison with any magnetic (exchange)
interaction. The conditions in oxides are such that small polarons and
bipolarons are the most probable quasiparticles both in insulating and
superconducting compounds. An opposite conclusion, recently suggested in the
literature \cite{nag,cha}, stems from (i) an incorrect expression for the
polaron binding energy, (ii) misunderstanding of  the criteria for small
polaron formation, and from (iii) misunderstanding  of the screening in
polaronic conductors. We have
developed the theory of the optical conductivity in doped magnetic
charge-transfer insulators with a strong electron-phonon interaction. The
spectral and temperature features of the optical conductivity of
ferromagnetic manganites are well described by the bipolaron absorption in
the paramagnetic phase and by the small polaron absorption in the
ferromagnetic phase. The pair breaking by exchange interaction with the
localized Mn spins explains the sudden spectral weight transfer in the
optical conductivity below $T_{c}$. Therefore, the optical probe of the
incoherent charge dynamics in manganites provides another strong evidence
for the carrier density collapse, which we proposed earlier as the
explanation of CMR. The anomalous specific heat and tunneling measurements also
support CCDC \cite{alebra2}.
The Mn localised spins give a major
contribution into magnetic susceptibility peak around $T_c$, and a
deviation of its temperature dependence from standard Curie behavior
may better elucidate a (smaller) magnetic response of polarons and a
character of the phase transition.

Our picture is further supported by recent ARPES data 
\cite{wsdan}, clearly showing a pseudogap in the band dispersion in
manganites at the Fermi level at low temperatures, incompatible with the
presence of a metallic phase. In addition, strong antiferromagnetic
correlations in the paramagnetic phase at $T>T_{c}$, found in polarized
neutron scattering and in Raman spectra\cite{wsaeppli} support our idea of
the singlet pairing above T$_{c}$. Outside the ferromagnetic regions (i) the
bipolarons would tend to form charge density waves, usually observed as a
charge ordering, and (ii) doped carries would spill over to Mn sites from
oxygen\cite{kannan}, thus allowing, at least in principle, a possibility for
some orbital effects. The theory suggests that by replacing the magnetic
ions (Mn) with nonmagnetic ions (Cu), one can transform a doped
charge-transfer insulator into a high-temperature superconductor owing to
the Bose-Einstein condensation of bipolarons \cite{alemot}, 
as was most likely observed experimentally\cite{bed}.

We are grateful to G. Aeppli, D.N. Basov, A.R. Bishop, D.S. Dessau, D.M.
Eagles, T. Egami, J.P. Franck, F. Fujimori, P. Horsch, M.F. Hundley, T.
Ishikawa, P.E. Kornilovitch, K.M. Krishnan, P.B. Littlewood, T.W. Noh,
A.P. Ramirez, G. A. Sawatzky, R.S. Williams, J. Zaanen, R. Zeyher, and
Guo-meng Zhao for helpful discussions. ASA acknowledges support from
the Quantum Structures Research Initiative and the External Research
Program of Hewlett-Packard Laboratories (Palo Alto).



\end{document}